\documentclass[aps,superscriptaddress]{revtex4}
\usepackage{hyperref,amssymb,graphicx,bbm}
\usepackage{amsfonts,amsmath,latexsym,times}
\usepackage {pgfplots}
\usepackage{tikz}

\begin{document}

\title{Beyond the Median Voter: A Model of How the Ideological Dimension Shapes Party Polarization}

\author{Roberto Venegeroles}
\email{roberto.venegeroles@ufabc.edu.br}
\affiliation{Centro de Matem\'atica, Computa\c c\~ao e Cogni\c c\~ao, Universidade Federal do ABC, 09210-170, Santo Andr\'e, SP, Brazil}

\date{\today}

\begin{abstract}
Defying the median voter theorem, party polarization has spread globally, especially in the United States. As concerns grow over its risks to democracy, political science has probed its causes, revealing two paradoxes: while polarization between U.S. political parties has undeniably increased since the 1970s, the corresponding rise in issue polarization among voters remains contested; moreover, the growing number of politically relevant issues would be expected to counteract party polarization rather than reinforce it. To examine these findings theoretically, we analyze a mathematical model of bipartisan elections where voters and parties interact in a multidimensional ideological space. We derive equations that determine the critical threshold of expected voter support needed for a party to find it strategically advantageous to position itself at the center or outside of it. This threshold fundamentally increases with the decrease of the effective dimension of the ideological space, while issue polarization among voters plays a secondary role. Thus, the model points to both paradoxes being explainable if ideological dimensionality is causally linked to polarization. The model reveals a Curie point: below this threshold, parties polarize as in a ferromagnetic phase, with voters acting like aligned spins; above it, they shift toward the center, akin to a paramagnetic phase.
\end{abstract}

\maketitle

\section{Introduction}
\label{sec1}

Political polarization is a global phenomenon \cite{COD} that has attracted intense interest not only from political scientists but also from the public. It is particularly pronounced in the U.S., where its evolution has been well documented \cite{COD,SMT}: From 1950 to 1974, the average Democrat remained separated from the average Republican by about 3 standard deviations in the House and 2.5 in the Senate \cite{SMT}. However, from 1974 to 2004, these differences increased to more than 5 standard deviations in the House and nearly 5 in the Senate \cite{SMT}. While in 1974, it was possible to classify 252 members of the House and 40 senators as falling between the most liberal Republican and the most conservative Democrat, by 2004, only four senators were in this overlap \cite{SMT}.

As political parties are representative entities, it is expected that party polarization would have some relationship with mass polarization of voters \cite{LMC}. However, while there is no doubt about the former, the latter remains a controversial subject in political science \cite{MM}. Some scholars understand the existence of mass polarization as an increase in the dispersion of issue distributions over time, which would represent a growing divergence in individuals' opinions \cite{AIA,BG,ALS,BYS}. Others conclude that such polarization does not exist, as they do not observe bimodality in these distributions, arguing that the majority of the population would be situated at the center of the ideological spectrum \cite{FJA,FAP,FAJP}. Paradoxically, from the perspective of issue-based ideology, the past decades reveal an intriguing contrast between sharp polarization in the U.S. Congress and the lack of a corresponding shift in the public \cite{YAKM}.

One challenge in measuring ideology is separating issue-based from identity-based ideology. The first refers to positions on topics such as taxes, healthcare, education, civil rights, environmental regulation, or climate change, among others, while the latter pertains to the emotional attachment to an ideological group. Ideological identities, such as ``liberal'' and ``conservative,'' no longer merely reflect a set of political issue positions. Instead, social attachments to these labels drive identity-based polarization within the U.S. electorate. These identities have become increasingly emotionally charged, shaped by psychological and social dynamics, including group affiliation and opposition to outgroups. This evolution complicates both the measurement and interpretation of ideological identities, but progress has been made in this domain \cite{LM1,LM2}.

Ideology is typically measured under the assumption that it can be reduced to a single dimension, such as a simple spectrum from left to right. However, this approach carries the risk of ignoring significant differences between various issues. This can lead to difficulties in capturing the true diversity of people's opinions, as a measure of ideological consistency may inappropriately group people who hold divergent views on some issues but align closely on others, treating them as if they shared a uniform political perspective. This is just one of the reasons why ideology is better understood as a multidimensional construct, with issues being grouped into areas \cite{MM,LMC, DHO,SK,STSH}. The other is that, if ideology is inherently multidimensional, understanding whether the issue space has shifted over time is essential for explaining the growth of polarization.

Over the past several decades, the scope of politics has expanded significantly due to increased education and access to information \cite{CW,GH}, leading people to engage with a greater number of issues \cite{MCZ,EM,JEA}. If the issue space is more diverse, candidates and parties need to form more flexible coalitions, as agreement on one issue does not necessarily imply agreement on all others. However, despite the expectation that such diversity would lead to a more negotiated and less binary politics, polarization has intensified. A possible explanation for this other polarization paradox is the reduction in the dimensionality of the issue space \cite{KLLT,JB}: Although more topics are being debated, opinions on them can be shown to be highly correlated due to identity-based ideology \cite{BG,DSM,KM,BSSS}, effectively resulting in only one or two issue dimensions \cite{STSH,TG}.  While some studies point to this reduction as a central factor in the polarization process \cite{DP,WA}, the relationship between dimensionality and polarization has remained merely correlational, with no consensus that one necessarily implies the other \cite{B3M}. However, a recent model has been proposed to investigate this relationship at the individual level, simulating how partisan bias shapes social interactions and learning \cite{KLLT}.

In this paper, we present a multidimensional version of the Hotelling-Downs model of bipartisan electoral competition \cite{Downs,HH}, inspired by the unidimensional model of \cite{YAKM}. Here each voter not only selects a candidate or party based on ideological proximity but also considers a satisfaction function that decreases as this distance increases. The multidimensional aspect of the model proves essential for understanding why issue polarization among the public appears as a secondary factor, as well as the seemingly counterintuitive impact of increasing issue diversity, which, under the narrowing of the ideological dimension, intensifies party polarization rather than containing it.  These findings have been independently documented in the literature, and the model aligns with them, despite there being no assumptions in the model that directly favor such relationships. Finally, we identify a Curie point in the model: below this threshold, parties exhibit polarized behavior akin to a ferromagnetic phase, with voters acting like aligned spins; above this point, parties converge toward the center, resembling a paramagnetic phase where voters behave as disordered spins.

\section{Mathematical Model}
\label{sec2}

Here we introduce a multidimensional model of a two-party election where: 1) Voters decide which party to support based on their own positions in the ideological space and how they perceive the rival parties' positions; 2) a party is more likely to be chosen by a voter if, in the voter’s view, it is closer to their own position; 3) knowing the distribution of voters and how the two parties are perceived by them, the rival parties strategically adjust their positions in the ideological space to maximize their vote gains; 4) the rival parties compete in the election on equal footing; 5) the ideological space is multidimensional. The proposed model aims to offer a mathematical framework for party positioning in response to voters' preferences, particularly within the context of polarization, rather than investigating potential causes of voter polarization, as opinion dynamics models typically do \cite{BSSS,DS,BLSSS,PT,LHAJS,OSPS,MHOZIJ,PRKI}.

Consider that each party $i$ adopts a position $u_i$ in the ideology space, initially considered as the real line. The model assumes that the probability $s_{i}(x,u_{i})$ that a voter positioned at $x$ is satisfied with party $i$ decreases to zero with the distance $|x-u_{i}|$, assuming the global maximum $s_{i}=1$ at $x=u_{i}$, i.e., when a voter is perfectly aligned with party $i$. This property is more realistic than the assumption of the classical Hotelling-Downs model \cite{Downs,HH}, which assumes that voters have a uniform preference for the ideologically closest party.

We consider the case in which two parties compete for voters. Here, index $i$ denotes either party $1$ or $2$, and $j$ denotes the rival party. The expected share of potential voters at position $x$ who vote for party $i$ is calculated according to \cite{YAKM}
\begin{eqnarray} 
p_{i}(x|u_{i}u_{j})=s_{i}(x,u_{i})\left[1-s_{j}(x,u_{j})\right]+\frac{1}{2}s_{i}(x,u_{i})s_{j}(x,u_{j}),
\label{pi}
\end{eqnarray}
where the first term on the right of Eq. (\ref{pi}) represents the expected share of voters at position $x$ who are satisfied with party $i$ only, and the second represents the equal sharing of those who are equally satisfied with both parties. In Eq. (\ref{pi}), it is also assumed that voters who are satisfied with none of the parties abstain from voting. A similar idea to $s_{i}(x,u_{i})$ as a ``preference function'' for party $i$ was previously proposed in \cite{HO}, but instead of Eq. (\ref{pi}), the expected values of each party's preference function (voting intentions) were defined by the midpoint $(u_{1}+u_{2})/2$, where the left party is chosen for positions up to the midpoint, and the right party for positions beyond it. Another alternative to (\ref{pi}) was to consider $p_{i} = U_{i} / \sum_{k} U_{k}$, where $U_{i}$ is the utility of party $i$, and abstention is treated as a type of candidate \cite{JSF}.

So far we have followed the steps of the ``satisficing'' model proposed in \cite{YAKM}, where $s_{i}(x,u_{i})$ is the satisficing function for party $i$. But here we will consider that the ideology space is multidimensional, with $x$, $u_{i}$ $\in \mathbb{R}^{n}$. Then the expected number of voters for party $i$, denoted by $V_{i}$, will be obtained by volume integration of (\ref{pi}) over the ideological space weighted by the population density $\rho(x)$
\begin{eqnarray} 
V_{i}(u_{i},u_{j})=\int p_{i}(x|u_{i}u_{j})\rho(x)d^{n}x.
\label{vi}
\end{eqnarray}
The population density $\rho(x)$ represents how voters are distributed across ideological space, and here we will assume that $\rho(x)$ is the multivariate normal mixture
\begin{eqnarray}
\rho(x)=\frac{|\Sigma_{0}|^{1/2}}{2(2\pi)^{n/2}}\left[e^{-\frac{1}{2}(x-d)^{T}\Sigma_{0}(x-d)}+e^{-\frac{1}{2}(x+d)^{T}\Sigma_{0}(x+d)}\right],
\label{rho}
\end{eqnarray}
where $d\in\mathbb{R}^{n}$, $\Sigma_{0}^{-1}$ is a $n\times n$ covariance matrix, and $|\Sigma_{0}|$ denotes the determinant of $\Sigma_{0}$. Note that, for convenience, and unlike the usual notation in the literature, $\Sigma_{0}$ is used here instead of $\Sigma_{0}^{-1}$. Mixture (\ref{rho}) indicates that voters cluster around the two poles of the ideological space, at \(x = -d\) on the left and \(x = +d\) on the right. Criteria for determining whether a mixture is unimodal or bimodal can be found in \cite{RGL}.

If party $i$ occupies the position $x=u_{i}$, we assume that $s_{i}$ follows a multivariate normal function given by
\begin{eqnarray} 
s_{i}(x,u_{i})=e^{-\frac{1}{2}(x-u_{i})^{T}\Sigma_{i}(x-u_{i})},
\label{si}
\end{eqnarray}
where $\Sigma_{i}$ is the $n\times n$ electoral preference matrix for party $i$, playing a role similar to that of the inverse of a covariance matrix, although $s_{i}$ is not a probability density function. It is important to highlight that each matrix $\Sigma_{i}$, like $\Sigma_{0}$, is positive-definite, ensuring that $s_{i}$ attains the global maximum at the party position $x=u_{i}$.

Before proceeding with the calculations, it is important to introduce two matrices that will be essential throughout this paper:
\begin{eqnarray} 
\Omega_{i}&=&\Sigma_{0}\left(\Sigma_{0}+\Sigma_{i}\right)^{-1}\Sigma_{0},
\label{ome}\\
\Phi&=&\Sigma_{0}\left(\Sigma_{0}+\Sigma_{1}+\Sigma_{2}\right)^{-1}\Sigma_{0},
\label{phi}
\end{eqnarray}
where the matrix $\Omega_{i}$ is associated with the voting intention $\langle s_{i}\rangle_{\rho}$ for party $i$, whereas $\Phi$ is associated with the joint voting intention $\langle s_{i}s_{j}\rangle_{\rho}$ for both parties. Now, the calculation of the expected number of voters for a party (\ref{vi}) through Eqs. (\ref{rho}) and (\ref{si}) is somewhat involved, but it suffices to mention that it involves integrals of the following form
\begin{eqnarray}
\int e^{-\frac{1}{2}x^{T}Ax+v^{T}x}d^{n}x=\sqrt{\frac{(2\pi)^{n}}{|A|}}e^{\frac{1}{2}v^{T}A^{-1}v},
\label{gauss}
\end{eqnarray}
which was reduced to a multivariate Gaussian normalization after the change of variables $x=y+A^{-1}v$, where $A$ denotes a matrix and $v$ denotes a vector. The final calculation yields the following result
\begin{eqnarray}
V_{i}=\frac{1}{2}\sqrt{\frac{|\Omega_{i}|}{|\Sigma_{0}|}}\frac{e^{\frac{1}{2}J_{i+}^{T}\Omega_{i}J_{i+}}+e^{\frac{1}{2}J_{i-}^{T}\Omega_{i}J_{i-}}}{e^{\frac{1}{2}\left(u_{i}^{T}\Sigma_{i}u_{i}+d^{T}\Sigma_{0}d\right)}}-\frac{1}{4}\sqrt{\frac{|\Phi|}{|\Sigma_{0}|}}\frac{e^{\frac{1}{2}K_{+}^{T}\Phi K_{+}}+e^{\frac{1}{2}K_{-}^{T}\Phi K_{-}}}{e^{\frac{1}{2}\left(u_{1}^{T}\Sigma_{1}u_{1}+u_{2}^{T}\Sigma_{2}u_{2}+d^{T}\Sigma_{0}d\right)}},
\label{vii}
\end{eqnarray}

where
\begin{eqnarray} 
J_{i\pm}&=&\Sigma_{0}^{-1}\Sigma_{i}u_{i}\pm d,
\label{jota}\\
K_{\pm}&=&\Sigma_{0}^{-1}\left(\Sigma_{i}u_{i}+\Sigma_{j}u_{j}\right)\pm d.
\label{ka}
\end{eqnarray}

Our focus here will be to investigate the dispute between two parties that are identical in strength, that is, $\Sigma_{1}=\Sigma_{2}=\Sigma$, and that will therefore be strategically aligned in symmetrical positions in the ideological space, i.e., $u_{1}=-u_{2}=u$. In this case, the expected number of voters $V_{i}=V$ will be
\begin{eqnarray}
V=\frac{1}{2}\sqrt{\frac{|\Omega|}{|\Sigma_{0}|}}\frac{e^{\frac{1}{2}J_{+}^{T}\Omega J_{+}}+e^{\frac{1}{2}J_{-}^{T}\Omega J_{-}}}{e^{\frac{1}{2}\left(u^{T}\Sigma u+d^{T}\Sigma_{0}d\right)}}-\frac{1}{2}\sqrt{\frac{|\Phi|}{|\Sigma_{0}|}}\frac{e^{\frac{1}{2}d^{T}\left(\Phi-\Sigma_{0}\right)d}}{e^{u^{T}\Sigma u}},
\label{vig}
\end{eqnarray}
where $\Omega_{i}=\Omega$ and $J_{i\pm}=J_{\pm}=\Sigma_{0}^{-1}\Sigma u\pm d$.

\section{Voting Threshold for Center Positioning}
\label{sec3}

Parties will position themselves strategically in the ideological space in order to maximize their voting potential. After performing the involved calculation of $\arg\max_{u_{i}} V(u_{i})$ for parties that are identical in strength, the positions $u_{1}=-u_{2}=u_{*}$ that maximize the total expected votes will be given by $u_{*}=0$ and, alternatively,
\begin{eqnarray}
u_{*}^{T}\left[\Sigma+\Sigma\left(\Sigma_{0}+\Sigma\right)^{-1}\Sigma\right]u_{*}&=&2\ln\left[\frac{\psi_{l}(u_{*})}{\beta}\right],
\label{opt}
\end{eqnarray}
for all $1\leq l\leq n$, where
\begin{eqnarray}
\label{bet}
\beta&=&2\sqrt{\frac{|\Omega|}{|\Phi|}}\,e^{-\frac{1}{2}d^{T}\left(\Phi-\Omega\right)d},\\
\psi_{l}(u)&=&\frac{\Sigma_{l}u}{\Sigma_{l}\left[\mathbb{I}-\left(\Sigma_{0}+\Sigma\right)^{-1}\Sigma\right]u\cosh\left(u^{T}\Sigma\Sigma_{0}^{-1}\Omega d\right)-\Sigma_{l}\Sigma_{0}^{-1}\Omega d\sinh\left(u^{T}\Sigma\Sigma_{0}^{-1}\Omega d\right)},
\label{psi}
\end{eqnarray}
and $\Sigma_{l}$ is the $l$th row of $\Sigma$. For the symmetric solutions $\pm u_{*}$ away from the center, the expected number of voters for a party will be given by substituting (\ref{opt}) into (\ref{vig}), which yields
\begin{eqnarray}
V_{*}=\frac{1}{2}\sqrt{\frac{|\Phi|}{|\Sigma_{0}|}}e^{\frac{1}{2}d^{T}\left(\Phi-\Sigma_{0}\right)d}e^{-u_{*}^{T}\Sigma u_{*}}\left[\psi_{l}(u_{*})\cosh\left(u_{*}^{T}\Sigma\Sigma_{0}^{-1}\Omega d\right)-1\right].
\label{V*}
\end{eqnarray}

Given that $\Sigma_{0}$ and $\Sigma$ are positive-definite matrices, the left-hand side of Eq. (\ref{opt}) is always positive for any $u_{*} \neq 0$. Consequently, equilibrium solutions outside the center only exist if $\psi_{l}(u_{*}) > \beta$. Thus, there is a critical transition between the strategic positioning of parties outside and at the center, which occurs when the two symmetrical positions $\pm u_{*}$ coalesce at the center. For this transition to occur, the following limit must exist independently of $l$
\begin{eqnarray}
\lim_{u\rightarrow0}\psi_{l}(u)=\beta_{c},
\label{lim}
\end{eqnarray}
where the critical parameter $\beta_{c}$ is such that, for $\beta<\beta_{c}$, the equilibrium occurs symmetrically outside the center $\pm u_{*}\neq0$, whereas for $\beta>\beta_{c}$, the center $u_{*}=0$ becomes the only equilibrium solution. To see this, it suffices to note from Eqs. (\ref{vig}) and (\ref{V*}) that
\begin{eqnarray}
\frac{V(u_{*}=0)}{V_{*}(u_{*}\rightarrow0)}=\frac{\beta-1}{\beta_{c}-1},
\label{rto}
\end{eqnarray}
where $\beta_{c}>1$ since $V_{*}>0$. Thus, for a dynamical modeling of the type $\dot{u} = \nabla_{u} V$, where $u^{*}=0$ and Eq. (\ref{opt}) represent equilibrium points, the transition along $\beta_{c}$ corresponds to a supercritical pitchfork bifurcation, similar to that observed in \cite{YAKM} in the one-dimensional case. It is also important to note that the quantities defined in Eqs. (\ref{ome}) and (\ref{phi}), through Eq. (\ref{bet}), crucially define the critical transition between the pursuit of the median voter and the onset of polarization.

The limit (\ref{lim}) exists only for a specific (critical) configuration $\Omega=\Omega_{c}$, which must be such that
\begin{eqnarray}
\Omega_{c}=\beta_{c}^{-1}\Sigma_{0}\left[\mathbb{I}+dd^{T}\left(\Omega_{c}-\Sigma_{0}\right)\right]^{-1}.
\label{sigc}
\end{eqnarray}
To calculate the expected total number of voters for a party in Eq. (\ref{vig}) at the critical point we need to calculate $|\Omega_{c}|$ and $|\Phi_{c}|$. In the case of $|\Omega_{c}|$, the determinant of Eq. (\ref{sigc}) can be obtained using the matrix determinant lemma \cite{KPM}, resulting in
\begin{eqnarray}
|\Omega_{c}|=\frac{|\Sigma_{0}|}{\beta_{c}^{n}\left[1+d^{T}\Omega_{c}\left(\mathbb{I}-dd^{T}\Sigma_{0}\right)^{-1}d\right]\left(1-d^{T}\Sigma_{0}d\right)}.
\label{xi1}
\end{eqnarray}
To determine $|\Phi_{c}|$, we first note that $\Phi^{-1}=2\Omega^{-1}-\Sigma_{0}^{-1}$, and then  proceed with similar calculations, resulting in
\begin{eqnarray}
|\Phi_{c}|=\frac{|\Sigma_{0}|}{(2\beta_{c}-1)^{n}\left[1+c_{\beta}d^{T}\Omega_{c}\left(\mathbb{I}-c_{\beta}dd^{T}\Sigma_{0}\right)^{-1}d\right]\left(1-c_{\beta}d^{T}\Sigma_{0}d\right)},
\label{zt1}
\end{eqnarray}
where $c_{\beta}=(2\beta_{c})/(2\beta_{c}-1)$. Another necessary result directly concerns the critical matrix $\Omega_{c}$ itself. Fortunately, obtaining $\Omega_{c}$ is not necessary, as this would require solving a second-order matrix equation, which is rarely solvable analytically. Instead, we need to determine the quantity below
\begin{eqnarray}
\omega_{c}=d^{T}\left(\Omega_{c}-\Sigma_{0}\right)d.
\label{omg}
\end{eqnarray}

Note that Eqs. (\ref{sigc}-\ref{omg}) still depend on the calculation of matrix inversions, which can be solved using the Sherman-Morrison formula \cite{KPM}, finally resulting in
\begin{eqnarray}
\omega_{c}&=&-\frac{1}{2}\left(1+d^{T}\Sigma_{0}d\right)+\frac{1}{2}\sqrt{\left(1-d^{T}\Sigma_{0}d\right)^{2}+\frac{4}{\beta_{c}}d^{T}\Sigma_{0}d},
\label{omef}\\
\frac{|\Omega_{c}|}{|\Phi_{c}|}&=&\frac{1+c_{\beta}\omega_{c}}{1+\omega_{c}}\left(2-\frac{1}{\beta_{c}}\right)^{n},
\label{rto}\\
\beta_{c}&=&2\sqrt{\frac{|\Omega_{c}|}{|\Phi_{c}|}}\exp\left\{\frac{\omega_{c}}{2}+\left[1-\frac{1}{(2\beta_{c}-1)(1+c_{\beta}\omega_{c})}\right]\frac{d^{T}\Sigma_{0}d}{2}\right\},
\label{betf}
\end{eqnarray}
where Eq. (\ref{betf}) follows from Eq. (\ref{bet}) at the critical point, from Eq. (\ref{omg}), and from the result to follow, which is derived using the same linear algebra relations that led to Eq. (\ref{omef}):
\begin{eqnarray}
d^{T}\Phi_{c}d=\frac{1}{2\beta_{c}-1}\frac{d^{T}\Sigma_{0}d}{1+c_{\beta}\omega_{c}}.
\label{dphi}
\end{eqnarray}

Finally, the critical value for the expected number of voters in a party is obtained from \(V(u=0)\) in Eq. (\ref{vig}) at the critical point, the identities (\ref{bet}) and (\ref{omg}), and Eq. (\ref{xi1}) applied to the Sherman-Morrison formula \cite{KPM}, resulting in
\begin{eqnarray}
V_{c}=\left(1-\frac{1}{\beta_{c}}\right)\sqrt{\frac{e^{\omega_{c}}}{\beta_{c}^{n}(1+\omega_{c})}}.
\label{vcf}
\end{eqnarray}
The output critical value $V_{c}$ in Eq. (\ref{vcf}) is the lowest expected number of voters for a party from which it is strategically more advantageous to remain at the center. Although Eqs. (\ref{opt}-\ref{psi}) are useful for determining the strategic positioning of rival parties in the ideological space, and thus the polarization $2u_{*}$ between them, they depend on the entries of $\Sigma_{0}$ and $\Sigma$, as well as the population separation $d$, totaling $n^{2}+n$ entries. On the other hand, the critical value $V_{c}$ will serve as our thermometer for party polarization, depending only (and explicitly) on the dimension of the ideological space $n$ and the square of the Mahalanobis distance $d^{T}\Sigma_{0}d$, via Eqs. (\ref{omef}) and (\ref{betf}), with $\beta_{c}$ and $\omega_{c}$ being intermediate parameters in these equations.

\section{Numerical results and limiting cases}
\label{sec4}

We can start by evaluating the unimodal population case $d=0$, for which Eqs. (\ref{omef})--(\ref{betf}) and (\ref{vcf}) reduce to
\begin{eqnarray}
\beta_{c}^{n+2}&=&4(2\beta_{c}-1)^{n},
\label{bet0}\\
V_{c}&=&\left(1-\frac{1}{\beta_{c}}\right)\frac{1}{\sqrt{\beta_{c}^{n}}}.
\label{V0}
\end{eqnarray}
In the case of a one-dimensional ideological space, with $\Sigma_{0}=\sigma_{0}^{-2}$ e $\Sigma=\sigma^{-2}$, Eq. (\ref{sigc}) reduces to $\sigma_{c}=\sigma_{0}(\beta_{c}-1)^{-1/2}$, meaning $\sigma_{c}/\sigma_{0}\approx0.807$, in agreement with \cite{YAKM}, see Table \ref{Tab1c} for $d=0$ and $n=1$. In this case, $V_{c}\approx0.38$, meaning that unless the ideological matrices $\Sigma_{0}$ for voter positions and $\Sigma$ for party preference grant each party at least $38\%$ of the electorate, these parties will not find strategic motivation to position themselves at the center of the ideological space. This number decreases to approximately $21\%$ in the case of a two-dimensional ideological space, and from $n=3$ onward, rival parties begin to strategically seek the ideological center, as occurs in the median voter theorem.

\begin{figure}[!htb]
    \centering
    \includegraphics[scale=0.5]{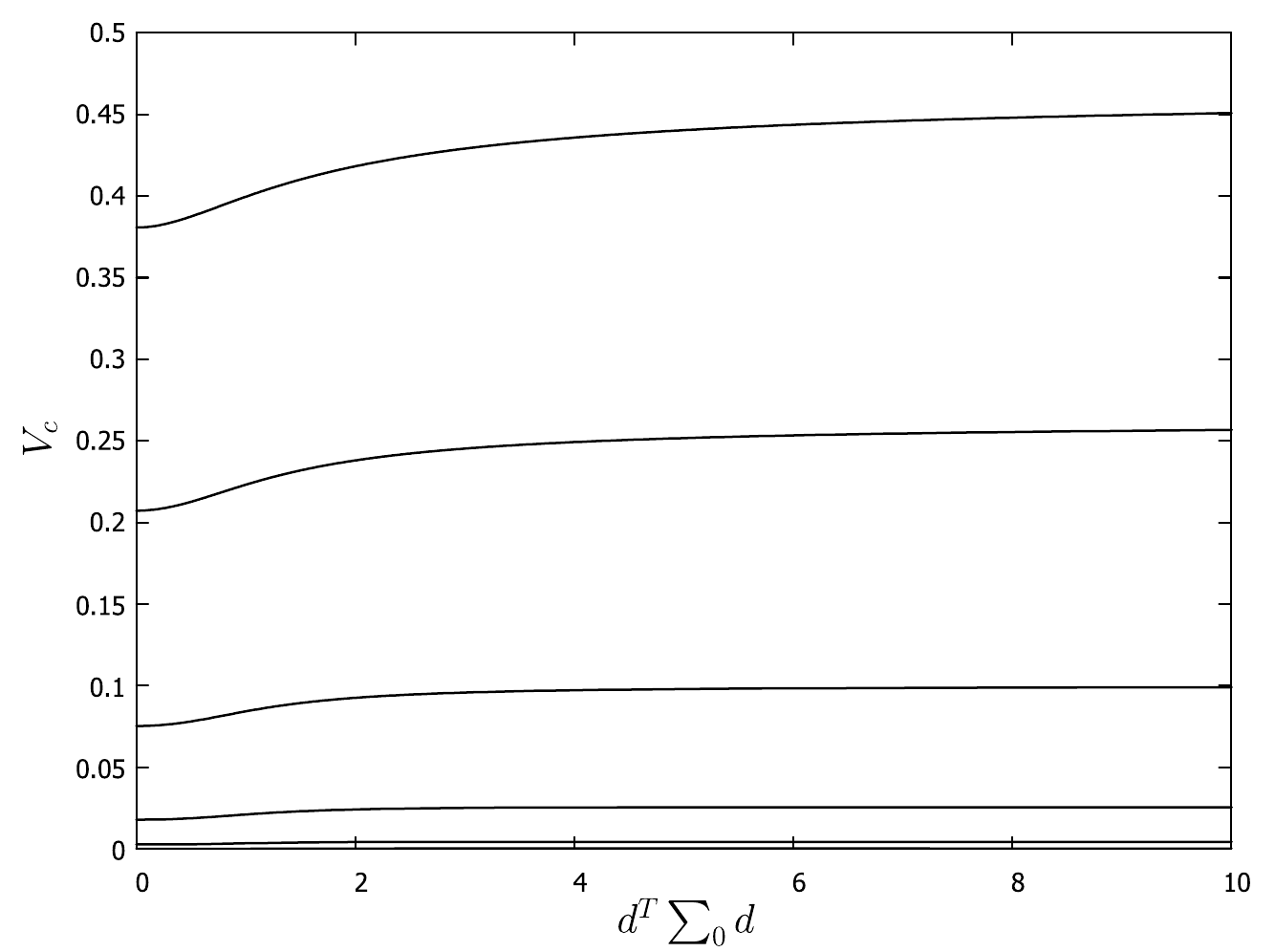}
    \caption{Reduction in the minimum expected number of voters for a party to strategically move to the center as the dimension of the ideological space $n$ increases, as a function of $d^{T}\Sigma_{0}d$: From top to bottom, the dimensions $n$ range from 1 to 5, consecutively.}
    \label{fig1}
  \end{figure}

In addition to Table \ref{Tab1c} , where we provide a more detailed description of how the critical parameters $\beta_{c}$ and $V_{c}$ evolve with the dimension $n$ and with the squared distance $d^{T}\Sigma_{0}d$, we also present in Fig. \ref{fig1} some graphs that allow for a broader visualization of these quantities. The graphs of $V_{c}$ in Fig. \ref{fig1} can be interpreted as the heights of vote barriers that a party must overcome to strategically pursue the ideological center in a two-party electoral contest. These barriers clearly decrease with each increase in the dimensionality of the ideological space. This indicates that as voters express a greater diversity of interests, rival parties will increasingly seek the median voter's support by adjusting their political agendas toward the center. Conversely, the less diverse the voters' concerns, the more polarized the parties will become in response. It is remarkable to observe from the graph how the issue polarization factor among voters, $d^{T} \Sigma_{0} d$, plays a secondary role in the growth of party polarization compared to the ideological space dimension $n$ of the competition.

Finally, Fig. \ref{fig1} suggests an asymptotic behavior of \(V_{c}\) as \(d^{T}\Sigma_{0}d \rightarrow \infty\). In fact, in this limit, \(\omega_{c}\) in Eq. (\ref{omef}) can be expanded as $\omega_{c} = 1 - \beta_{c}^{-1} + O(1/d^{T}\Sigma_{0}d)$, and the equations for \(\beta_{c}\) and \(V_{c}\) in this case reduce to
\begin{eqnarray}
\beta_{c}^{n+1}&=&4(2\beta_{c}-1)^{n-1}e^{1-1/\beta_{c}},\\
\label{binf}
V_{c}&=&\left(1-\frac{1}{\beta_{c}}\right)\sqrt{\frac{e^{-1+1/\beta_{c}}}{\beta_{c}^{n-1}}},
\label{vinf}
\end{eqnarray}
with some asymptotic values of $\beta_{c}$ and $V_{c}$ displayed in the last columns of Table \ref{Tab1c}.

\begin{table}
  \begin{center}
    \scriptsize
    \setlength{\tabcolsep}{6pt}
    \renewcommand{\arraystretch}{1.1}
    \caption{Dependence of the critical expected number of voters for a party on the ideological dimension $n$ for certain values of $d^{T}\Sigma_{0}d$.}
    \vspace{2mm}
    \begin{tabular}[c]{c|c|c|c|c|c|c|c|c|c|c}
    \hline
 &\multicolumn{2}{l|}{\,\,\,\,\,\,\,\,\qquad$d=0$}&\multicolumn{2}{l|}{\,\,\,\,\,\qquad$d^{T}\Sigma_{0}d=1$}&\multicolumn{2}{l|}{\,\,\,\,\,\qquad$d^{T}\Sigma_{0}d=5$}&\multicolumn{2}{l|}{\,\,\,\,\,\qquad$d^{T}\Sigma_{0}d=10$}&\multicolumn{2}{l|}{\,\,\,\,\,\qquad$d^{T}\Sigma_{0}d\rightarrow\infty$}\\
      \hline\hline
      $n$ & $\beta_{c}$ & $V_{c}$ & $\beta_{c}$ & $V_{c}$ & $\beta_{c}$ & $V_{c}$ & $\beta_{c}$ & $V_{c}$ & $\beta_{c}$ & $V_{c}$\\\hline
      1&  $2.53407$ & 0.380292 & $2.55299$ & $0.399129$ & $2.65222$ & $0.439868$ & $2.69218$ & $0.450299$ & $2.74909$ & $0.462876$\\\hline
      2&  $3.41421$ & 0.207107 & $3.44110$ & $0.222986$ & $3.62610$ & $0.251505$ & $3.70299$ & $0.256454$ & $3.81249$ & $0.261269$\\\hline
      3&  $4.79564$ & 0.075365 & $4.82958$ & $0.084337$ & $5.14486$ & $0.097990$ & $5.27862$ & $0.099045$ & $5.46830$ & $0.099312$\\\hline
     4&  $6.87936$ & 0.018059 & $6.91884$ & $0.021259$ & $7.42229$ & $0.025675$ & $7.63860$ & $0.025647$ & $7.94398$ & $0.025217$\\\hline
     5&  $9.94482$ & 0.002884 & $9.98850$ & $0.003605$ & $10.7595$ & $0.004560$ & $11.0933$ & $0.004495$ & $11.56275$ & $0.004327$\\\hline
     6&  $14.3849$ & 0.000312 & $14.4363$ & $0.000417$ & $15.5861$ & $0.000556$ & $16.0862$ & $0.000540$ & $16.7876$ & $0.000509$\\\hline
    \end{tabular}
    \label{Tab1c}
  \end{center}
\end{table}

\section{Final Remarks}
\label{sec5}

As the values of \( d^{T}\Sigma_{0}d \) increase, the bimodality of the voter distribution in ideological space becomes more pronounced, leading to more ideologically polarized voters in terms of issue positions. In the one-dimensional case, this results in increasingly higher values of $V_{c}$, promoting (moderate) a tendency for rival parties to be more likely to shift toward the extremes rather than converge toward the center, unless they overcome the growing $V_{c}$ barrier. This scenario was intuitively predicted by Downs \cite{Downs} in his considerations on Hotelling's one-dimensional voter model \cite{HH}. It is worth reproducing here Downs' reflections (1957) on the consequences of this scenario in light of current concerns:

\begin{quote}
``{\it This reasoning implies that stable government in a two-party democracy requires a distribution of voters roughly approximating a normal curve. When such a distribution exists, the two parties come to resemble each other closely. Thus, when one replaces the other in office, no drastic policy changes occur, and most voters are located relatively close to the incumbent's position, no matter which party is in power. But when the electorate is polarized, (...) \textnormal{[bimodal]}, a change in parties causes a radical alteration in policy. And, regardless of which party is in office, half the electorate always feels that the other half is imposing policies upon it that are strongly repugnant to it. In this situation, if one party keeps getting re-elected, the disgruntled supporters of the other party will probably revolt; whereas if the two parties alternate in office, social chaos occurs because government policy keeps changing from one extreme to the other. Thus, democracy does not lead to effective, stable government when the electorate is polarized. Either the distribution must change or democracy will be replaced by tyranny, in which one extreme imposes its will upon the other.}'' \cite{Downs}.
\end{quote}

Regarding party positioning strategies under bimodal electorates in one dimension, the results presented here align with Downs' findings, particularly his projections about the instability of democracy in such situations, which remain highly relevant in contemporary times \cite{JB,MS}. However, when considering electoral contests within a multidimensional ideological space, the stability of a two-party democracy does not necessarily require a unimodal distribution of voters. The results suggest that, at least in theory, if the ideological space is expanded (or perhaps restored), rival parties would find significantly more electoral motivation to seek the center as the vote barrier $V_{c}$ to be surpassed is significantly lower. Thus, even with a bimodal distribution of the electorate, parties would begin to resemble each other more closely, with no drastic policy changes regardless of which party is in power.

Clearly, the multidimensionality of the ideological space is not in question here; rather, it is the extent to which voters engage with these dimensions that is under consideration. Our results definitively confirm the causal connection between polarization and dimensionality reduction, settling the question raised in \cite{KLLT}. The effect of the dimensionality can be heuristically understood by considering that individuals occupying positions with the same sign across all issue axes can be classified as ideologically more ``purists.'' In an ideological space of dimension $n$ where all quadrants are significantly populated, the proportion of purists decreases with $n$ according to $2^{1-n}$. This dynamic incentivizes political parties to focus on voters closer to the median as the dimensionality $n$ increases. Naturally, the contraction of purists alone may not suffice to prompt an immediate partisan shift toward the median voter, since the voter distribution may still feature a significant concentration of ``nearly purist'' individuals (one axis of one sign, $n-1$ of the other, and so on). However, as the dimensionality increases, even these nearly purist voters gradually come to constitute a smaller share of the electorate. Consequently, the exercise of dimensionality, which represents greater plurality, acts as a moderating factor that potentially curtails partisan extremism, a counterbalance coming from individuals with ``cross-pressures'' \cite{LM}.

Downs \cite{Downs} also points to abstention as a possible brake on the tendency of candidates to converge to the median in the case of a bimodal distribution of voters, which can be understood as the effect of both the alienation of extremist and the indifference of centrist voters. The model presented here also takes into account dissatisfaction with both candidates in a two-party contest without any prior hypothesis regarding the distribution of voters, given by $(1-s_{i})(1-s_{j})$. Thus, when observing the values of $V_{c}$ and comparing them with the actual performance of candidates in U.S. elections, for example, it is also necessary to consider factors such as abstentions, votes for minor parties and independent candidates, and residual votes. These factors can act as obstacles to overcoming the critical barrier $V_{c}$, possibly serving as elements that, even if not sufficient as one of the explanations for growing party polarization \cite{MPF}, at least help to understand why the median voter theorem is hardly observed in moderate electoral scenarios. Additionally, the number and nature of issues often change from election to election, making it difficult to predict which ones will be relevant in future contests. Therefore, for empirical analysis, the issues, dimensions, and other parameters of a multidimensional model must be measured separately for each election \cite{DHO}.

Finally, the polarization model discussed here also presents interesting parallels with ferromagnetism. It is well known that the magnetization $m$ is the order parameter of a simple uniaxial ferromagnet, which exhibits a disordered phase above a critical temperature $T_{c}$, for which $m=0$, and an ordered phase below $T_{c}$, where $m$ can take nonzero symmetric values. This suggests that the position $u$ of a party plays a similar role as the order parameter in the polarization model, where $\beta$ would play the role of temperature. In the theory of phase transitions, Landau's expansion of the free energy $F(m,T)$ near the critical temperature is given by \cite{SS}
 \begin{eqnarray}
F(m,T)=-Hm+\frac{1}{2}a(T)m^{2}+\frac{1}{4}b(T)m^{4}+\ldots,
\label{land}
\end{eqnarray}
where $H$ is the external magnetic field, and $a$ and $b>0$ are temperature-dependent coefficients of the expansion, in particular $a(T)\propto(T-T_{c})$ near $T_{c}$. We can model dissipative dynamics for this ferromagnet through the phenomenological equation
 \begin{eqnarray}
\frac{dm}{dt}=-\gamma\frac{\partial F}{\partial m},
\label{diss}
\end{eqnarray}
with $\gamma>0$, causing the magnetization to evolve toward a local minimum of the free energy, which will be a decreasing function of time. After rescaling
 \begin{eqnarray}
m=\frac{u}{\sqrt{b\gamma}},\qquad \nu(\beta-\beta_{c})=\gamma a,
\label{resc}
\end{eqnarray}
with $\nu>0$ and $h=\sqrt{b\gamma^{3}}H$, one obtains
\begin{eqnarray}
\frac{du}{dt}&=&h-\nu(\beta-\beta_{c})u-u^{3}\equiv-\frac{\partial f}{\partial u},
\label{upt}\\
f(u,\beta, h)&=&-hu+\frac{\nu}{2}(\beta-\beta_{c})u^{2}+\frac{1}{4}u^{4}.
\label{free}
\end{eqnarray}
The normal form (\ref{upt}) is known as an imperfect bifurcation \cite{PM}, where $h$ determines the selection of one of the stable points below the critical parameter through a spontaneous symmetry breaking. The case of zero external magnetic field $h=0$ addresses the supercritical pitchfork bifurcation of the one-dimensional phenomenological modeling
 \begin{eqnarray}
\frac{du}{dt}=k\frac{\partial V}{\partial u},
\label{diss}
\end{eqnarray}
with $k>0$, as studied in \cite{YAKM}, and therefore $f(u,\beta, h=0)$ represents the corresponding free energy of the model. The parameter $\nu$ is such that $u_{*}=\pm\sqrt{\nu(\beta_{c}-\beta)}$ represent the equilibrium positions below the Curie point $\beta_{c}$, with exponent $1/2$ for the order parameter, while $u_{*}=0$ represents the equilibrium above $\beta_{c}$.

As in a ferromagnet, where the total magnetization results from the sum of individual magnetic moments of spins, the electoral landscape in our model emerges from the aggregation of individual voter choices. When ideological dimensionality is low, partisan alignment among voters mirrors the cooperative behavior of spins in a ferromagnetic phase, collectively reinforcing party polarization. Conversely, when the effective ideological dimension increases beyond a critical threshold, this alignment weakens, resembling the disordered state of a paramagnet, where voter preferences become more fluid and less constrained by partisan identities.

\section*{Acknowledgements}
The author thanks Alberto Saa (Unicamp) and Marcus Andr\'e Melo (UFPE) for helpful discussions.

\end{document}